\documentclass[a4paper,fleqn,usenatbib]{mnras}
\usepackage{newtxtext,newtxmath}
\usepackage[T1]{fontenc}
\usepackage{ae,aecompl}
\usepackage{graphicx}	% Including figure files
\usepackage{amsmath}	% Advanced maths commands
\usepackage{hyperref}
\LARGE \normalsize \title[{The black hole mass in MAXI
  J1659-152}]{Delimiting the black hole
  mass in the X-ray transient MAXI J1659-152 with H$\alpha$
  spectroscopy}
\author[Torres et al.]  {M.A.P.~Torres$^{1,2}$\thanks{email :
    mapt@iac.es}, P.G.~Jonker$^{3,4}$, J. Casares$^{1,2}$,
 J. C. A. Miller-Jones$^{5}$,  D.~Steeghs$^{6}$ \\
$^1$Instituto de Astrof\'{i}sica de Canarias, E-38205 La Laguna, S/C de Tenerife, Spain \\
$^2$Departamento de Astrof\'{i}sica, Universidad de La Laguna, E-38206
La Laguna, S/C de Tenerife, Spain \\
$^3$Department of Astrophysics/ IMAPP, Radboud University, P.O.~Box
9010, 6500~GL, Nijmegen, The Netherlands \\
$^4$SRON, Netherlands Institute for Space Research, Sorbonnelaan 2,
3584~CA, Utrecht, The Netherlands\\ 
$^5$International Centre for Radio Astronomy Research, Curtin University, GPO Box U1987, Perth, WA 6845, Australia \\
$^6$Department of Physics, University of Warwick, Coventry CV4 7AL, UK \\
}

\begin{document}

\maketitle

\begin{abstract} \noindent 

MAXI J1659-152 is a 2.4 h orbital period X-ray dipping transient  black hole
candidate. We present
spectroscopy of its $I\approx23$ quiescent counterpart where we
detect H$\alpha$ emission with full-width at half-maximum (FWHM) of $3200 \pm 300$ km
s$^{-1}$. Applying the correlation between the
H$\alpha$ FWHM and radial velocity semi-amplitude of the donor star
for quiescent X-ray transients, we
derive $K_2 = 750 \pm 80$ km s$^{-1}$. The orbital period and $K_2$
lead to a mass function of $4.4\pm1.4~M{_\odot}$ (1$\sigma$). The donor to compact object mass ratio and binary
inclination are likely in the range $q={M_2}/{M_1} = 0.02-0.07$
and $i=70{^\circ}-80{^\circ}$. These constraints imply a 68\%
confidence level interval for the compact object mass of  $3.3
\lesssim M_1(M_\odot) \lesssim 7.5$, confirming its black hole
nature. These quasi-dynamical limits are compared to mass estimates
from modelling of X-ray data and any discrepancies are
discussed. We review the properties of optical spectroscopy and time-series photometry collected
 during the 2010-2011 outburst. We interpret the apparent
 modulations found soon after  the onset of high-accretion activity and during the 2011
rebrightening event as originating in the accretion disc. These have
signatures consistent with superhumps, with the 2011 modulation having a fractional period excess
$< 0.6\%$ (3$\sigma$). We propose that direct irradiation of the
donor by the central X-ray source was not possible due to its
occultation by the disc outer regions. We argue that disc shielding
significantly weakens the donor star contribution to the optical
variability in systems with $q\lesssim0.07$, including  neutron star ultra-compact
X-ray binaries. 
\end{abstract}


\section{Introduction}

Black hole (BH) low-mass X-ray binaries accrete mass from a low-mass
Roche-lobe filling star. In these systems conservation of angular momentum dictates
that the accretion process proceeds through a disc.
Characteristically, they  show unpredictable transient episodes
of increased bolometric luminosity caused by enhanced disc accretion
onto the compact object. These episodes are  followed by long
periods of low accretion activity.  As X-ray transients they can evolve through 
distinct  states during outburst. These are phenomenologically defined
by the strength of the different components that make up the energy spectrum, 
the rms variability as measured in the
power density spectra, and the presence of specific types of quasi-periodic
oscillations (QPOs) - see e.g., \citet{2006ARAA..44...49R} for a
review. Obtaining dynamical BH masses and orbital parameters is key to
test accretion and ejection models that will eventually put on a firm footing
the processes responsible for the different spectral and time variability
observed in these systems (e.g., \citealt{2019ApJ...887...21S,2013ApJ...762..104S}; section \ref{sec:mass} in this work).

During the intervals of quenched X-ray luminosity, photospheric lines
from the donor star may be detected on top of the remaining
accretion-powered emission. In this favourable case, the X-ray transient can be studied at optical/infrared wavelengths as a
single-lined spectroscopic binary with a circular orbit where the
compact object mass function $f(M)$  is
established from the orbital period $P_{orb}$ and the radial velocity
semi-amplitude of the donor star $K_2$ as $f(M) = \frac{{K_2}^3
  P_{orb}}{2 \pi G}$. Mass functions exceeding 3 M$\odot$ unambiguously confirm  the BH nature of the compact object.  The
actual BH and donor masses can be obtained when the donor star to BH mass ratio
$q$ and  binary inclination $i$ are measurable. In this instance and
following  Kepler's laws:
$M_1= f(M) \left( 1+q \right)^2 \sin^{-3} i=  {M_2}/q$ where
$M_1$ and $M_2$ stand for the BH and donor star mass, respectively.
By virtue of tidal locking of orbit and stellar rotation, $q$ can be derived by measuring $K_2$ and the projected rotational
velocity of the donor star (see e.g., \citealt{2020ApJ...893L..37T} for a concise
explanation). To measure both quantities, the spectroscopic
observations have to reach a resolution, orbital coverage and signal
level above the noise that allow for accurate evaluation of the
radial velocity curve and line Doppler broadening engraved on the
photospheric features by the donor's orbital motion and rotation. 
And critically, for systems lacking eclipses, $i$ must be firmly 
established via
light curve modelling. This must disentangle the orbital phase-dependent
photometric modulation ascribed to the
distorted shape of the
donor star from any other  accompanying variability.  With the dynamical
method,  stellar-mass BHs have been measured or confirmed to date in
18 Galactic X-ray transients
(\citealt{2014SSRv..183..223C,2017ApJ...846..132H,2019ApJ...882L..21T}).  These BH mass determinations together with
knowledge of the Galactic distribution and kinematics of the host
binaries can serve as a common thread for inferring their formation
mechanisms (e.g., \citealt{2012ApJ...757...91B,2017MNRAS.467..298R,2019MNRAS.489.3116A,2020MNRAS.496L..22G}), exploring their evolutionary pathways as close
binaries \citep{2003MNRAS.341..385P,1999ApJ...521..723K} and
understanding the origin  of merging BHs identified with gravitational wave signals  \citep{2017PhRvD..95l4046G,2018arXiv180605820M}. 

For a large fraction of the candidate BH transients detected to date, the observational requirements for a
  dynamical study demand excessive use of telescope time under
  excellent image quality conditions or they are just not
  reachable. The former difficulty  is owned to the faintness of  the
  optical/infrared  counterparts
  (e.g., \citealt{2012MNRAS.423.2656R,2019MNRAS.482.2149L}). The latter impossibility
  can be due to the presence of  interloper field stars
(e.g., \citealt{2014MNRAS.439.1381R}) or the lack of donor  features \citep{2015MNRAS.450.4292T}. For
 these elusive sources alternatives have to be considered to constrain
 the system parameters. The first relies on time-series observations
 during an outburst event when the multi-wavelength counterparts to the
 X-ray transient become brighter. In well-monitored outbursts it is
 not unusual to  identify signatures of the donor star and BH  motion
 encoded in periodic modulations of the photometry, emission lines or both (see
    for instance \citealt{2006csxs.book..215C} for a
    review). More recent techniques exploit  simple-to-measure properties
    of the H$\alpha$ emission line profile emitted by the quiescent accretion  disc
    to reliably estimate or constrain $K_2$ and $q$
    \citep{2015ApJ...808...80C,
      2016ApJ...822...99C,2018MNRAS.473.5195C,2018MNRAS.481.4372C}. In
    this paper we apply several of these indirect techniques to optical data of the
    X-ray  transient MAXI J1659-152 acquired  during quiescence and
    outburst. We will confirm the BH nature of the compact
    object by setting solid constraints on its mass. 

MAXI J1659-152 (hereafter J1659) was first discovered at $\gamma$-ray
energies on 25 Sep 2010, 08:05:05 UT
by the {\it Swift} Burst Alert Telescope \citep{2010GCN.11296....1M}
and at X-ray wavelengths by the Monitor of All-sky X-ray Image Gas Slit
Camera  \citep{2010ATel.2873....1N}. The momentary misclassification
of the transient source with a $\gamma$-ray burst (GRB 100925A) triggered rapid
multi-wavelength follow-up observations. Thus, on the same date, a
16.8 mag optical counterpart was identified with  {\it Swift}/UVOT
\citep{2010GCN.11298....1M} and a spectrum taken with the Very Large
Telescope (VLT) revealed broad
emission features at zero redshift confirming the
Galactic X-ray binary nature of the transient
\citep{2010GCN.11307....1D}.  The main outburst event lasted
${\sim}220$ d \citep{2012MNRAS.423.3308J,2013ApJ...775....9H} during
which J1659 underwent X-ray variability and state changes
characteristic of confirmed accreting stellar-mass BHs
\citep{2011ApJ...731L...2K,2011ApJ...736...22K,2011MNRAS.415..292M}. The
analysis of the X-ray data collected during the outburst has
delivered %vast
estimates for the compact object mass
that span a large range from 2 to 20 M$\odot$
\citep{
2011arXiv1103.0531S,
2011ApJ...736...22K,
2012PASJ...64...32Y,
2013MNRAS.436.2625V,
2016MNRAS.460.3163M}.
Furthermore, \citet{2020ApJ...888L..30R} claim that the angular momentum vector of
the BH spin and orbital motion have opposite signs (e.g., a retrograde spinning BH).

During the early phase of the outburst, 
the X-ray light curve displayed recurrent dips revealing a high
inclination system and a potential 2.414 h orbital period
\citep{2010ATel.2877....1K,2010ATel.2912....1K,2010ATel.2926....1B,2013A&A...552A..32K}. The same periodicity was found in optical light curves obtained soon
after the discovery of the transient
\citep{2010fym..confP...4K,2010fym..confP...6O} and during a short (${\sim}80$
d long) rebrightening event that occurred at the end of the main outburst
\citep{2018MNRAS.475.1036C}. The optical counterpart seemed to
level off at $r{\sim}23.7$ in early (March 2012) X-ray quiescence
\citep{2012ApJ...760L..27K}. However, it further declined in
brightness being found at $r'=24.4$ by May 2014 and  $i'\approx23$ (in
June 2013/2015), showing aperiodic photometric  variability
\citep{2018MNRAS.475.1036C}.  The distance towards J1659 is constrained to
$8.6 \pm 3.7$ kpc and its height above the Galactic plane to $2.4 \pm
1.0$ kpc (although see \citealt{2013A&A...552A..32K}  for the assumptions used for these and early estimates).  The faintness of the quiescent optical counterpart
and short orbital period make J1659 a very challenging
target for a dynamical study. In this work we tackle the
current inability of obtaining both the radial velocity curve and
projected rotation velocity of the donor star, measurables that 
would otherwise allow to determine $K_2$  and $q$. We
will constrain these two quantities by characterizing the H$\alpha$
emission line detected during quiescence and by interpreting the
photometric modulations observed during outburst.

The manuscript is structured as follows: we start by
presenting in Section \ref{sec:obs} the optical spectroscopy and
the steps performed for their reduction and the extraction of the
H$\alpha$ emission line. We portray the properties of the
Balmer line profile in Section \ref{sec:ana} and we derive $K_2$ and limits on $q$. In this
Section we also review the origin of the
optical modulations observed during the 2010-2011
outburst and constrain $q$ for a superhump scenario. In Section
\ref{sec:dis} the BH mass is quantified and compared to estimates
provided by modeling of X-ray data. In addition, we discuss how in systems with low $q$ shielding of the central X-ray source by the outer disc regions is significant, reducing the heating of the donor star and hence its contribution to the optical variability.

\section{Observations and data reduction}
\label{sec:obs}
We obtained spectroscopy of  J1659 acquired in service mode 
at the Paranal observatory (Chile) of the European Southern
Observatory (ESO) under program number 091.D-0865(A).  The observations were performed with the
FOcal Reducer and low dispersion Spectrograph 2 (FORS2) which was
attached to the Cassegrain focus of the 8.2-m Unit 1 VLT. The instrument was used with the standard
resolution collimator and the two $2048 \times 4096$ pixels MIT CCDs
binned 2 by 2. This provided a 0\arcsec\!.25
 spatial scale per binned pixel.  The 300 line mm$^{-1}$ grism GRIS\_300I+11 and a
1\arcsec\!.0 wide slit were used, yielding a dispersion of 3.18 \AA~
pixel$^{-1}$, a  nominal $\lambda\lambda6020-11000$ \AA~ coverage and a
slit width-limited resolution of ${\sim}11$ \AA~FWHM. 

Four contiguous 620s spectra  of J1659 were collected on 6 June 2013
from 6:25 -
7:09 UT at airmass $1.11-1.23$ under a moonless sky. The brightness of the optical  counterpart was $I=23.3 \pm 0.1$ as
established from a photometric calibration of the  acquisition images \citep{2018MNRAS.475.1036C}\footnote{A finding chart built from the FORS2 acquisition imaging
is made available in the on-line version of BlackCAT
\citep{2016A&A...587A..61C}. Complementary charts are published in
\citet{2012ApJ...760L..27K}.}. By examining the width of the spatial profile
of the spectra of two field stars centered on the slit, we establish an image quality
of ${\sim}0\arcsec\!.7$ FWHM at wavelengths covering H$\alpha$.
Thus, a seeing-limited spectral resolution of ${\sim}7.9$ \AA~FWHM (= 360
km s$^{-1}$) was achieved. 

The data reduction was performed following standard procedures
implemented in {\sc iraf}. These consisted of de-biasing and
flat-fielding the science and calibration arc-lamp frames.  The latter
were obtained at the end of the night to determinate the
pixel-to-wavelength conversion in the spectra. The target 1-D spectra
were obtained using the {\sc kpnoslit} package. Given the very weak
optical continuum from J1659, the spectral profiles were not traced to
avoid any large departures from the target location during the
extraction. From examination of the spectrum of an off-axis bright
field star, the spatial curvature of the trace for J1659 along the
dispersion direction is estimated to be < 1.5 pixels in amplitude. After
several tests, an extraction aperture of size 6 pixels was chosen and
fixed at the position of the target spatial profile at H$\alpha$. To
obtain this profile, we first identified the locus of H$\alpha$ in the
reduced data utilizing sky emission lines as a wavelength map. Next,
we median combined the pixels along the dispersion axis that covered
the Balmer emission from J1659. By doing this we generate for each
spectrum the best possible target spatial profile, which is fit to
obtain its central position for the extraction.  For the sky
background determination, care was taken in selecting regions free of
emission from nearby field stars that fall on the slit.  The
subsequent pixel-to-wavelength calibration was derived through two-piece cubic spline
fits to 18 arc lines, displaying an rms scatter $< 0.12$ \AA.  The
resulting spectra of J1659 were imported to {\sc molly}, resampled into a
common uniform 145.6 km s$^{-1}$ pixel$^{-1}$ scale and shifted into the heliocentric
velocity frame. Finally, the wavelength zeropoint
was corrected employing a velocity shift of 47 km s$^{-1}$. This
offset was determined by comparing the
fitted central wavelength of the [O{\sc i}] $\lambda 6300.3$ sky line
(at the observer's restframe) with the actual wavelength.

The extracted individual 1-D spectra have a signal-to-noise ratio per
pixel of  ${\sim}1$ in regions of the continuum bracketing the H$\alpha$
emission (reaching ${\sim}3-4$ at maximum intensity of the line) and $<2
$ in regions covering the Ca {\sc ii} triplet.

\section{Data analysis and Results}
\label{sec:ana}

In this section we will employ the ephemeris for the mid-time of the
X-ray dip episodes observed during outburst and given in \citet{2013A&A...552A..32K}:  
$$Mid-dip = MJD~55467.0904 \pm 0.0005 + N \times (0.10058 \pm 0.000022)$$ 
\noindent where $N$ is the number of orbital cycles. This ephemeris
supersedes that initially communicated by \citet{2010ATel.2926....1B}.
The absorption dip events were regularly observed during the first
${\sim}10$ days since the outburst discovery on MJD~55464.10. Their
duty cycle, defined as the duration of the dip as a fraction of the
orbital period, was $\lesssim0.2$. 
We refer the reader to the $0.3 - 10$ keV light curve in fig.~2 of
Kuulkers et al.~to see the dipping activity contemporaneous to the
phenomenology described in this section. Finally, in this work we will quote fundamental parameters for X-ray binaries of interest citing the original work or, in the case of BH transients, the most reliable values presented in \citet{2014SSRv..183..223C} and \citet{2016ApJ...822...99C}. 

\subsection{The H$\alpha$ line profile}
\label{sec:ha}

\begin{figure}
\includegraphics[width=3.6in]{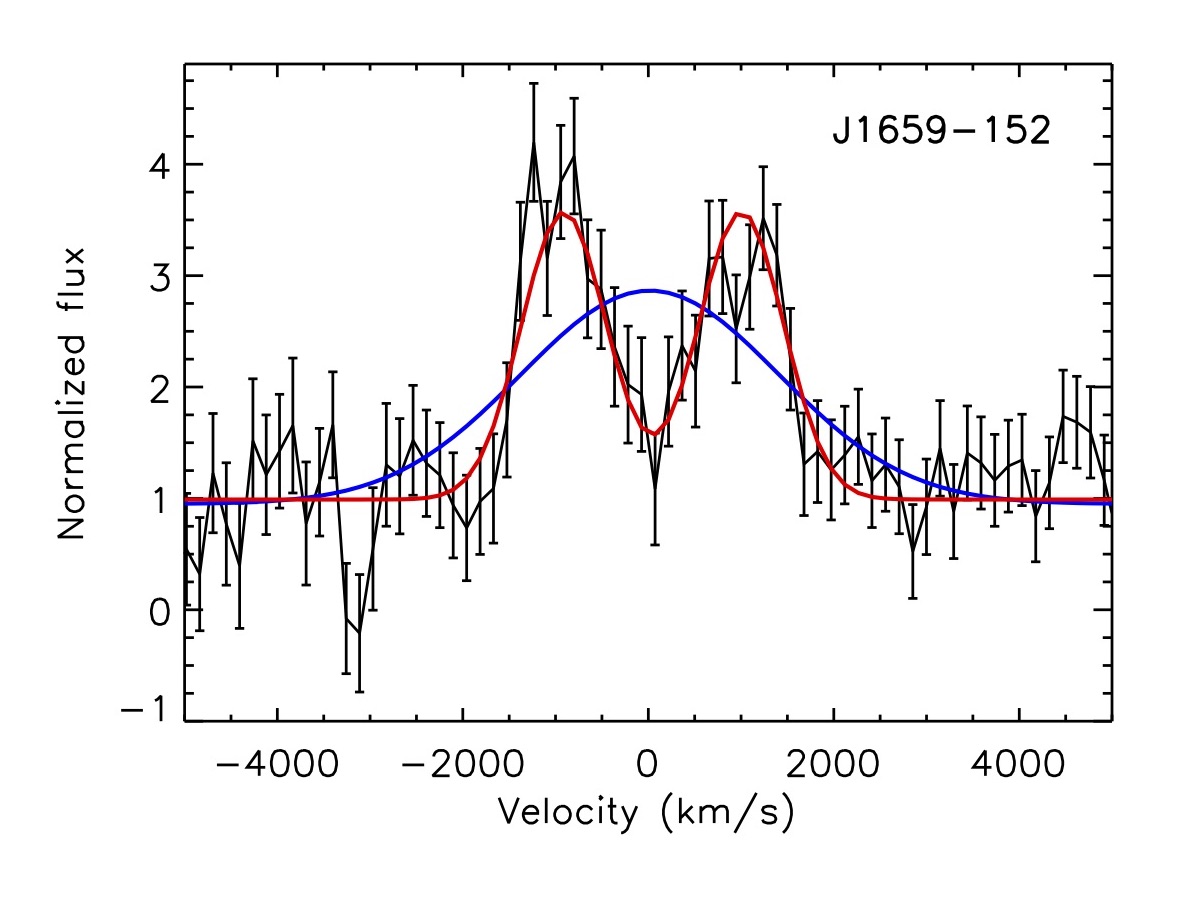}
\caption{Averaged normalized H$\alpha$ emission line
of J1659 detected during quiescence with associated
 uncertainties overplotted. The single and 2-Gaussian component fits
performed in Section \ref{sec:ha}  are displayed with blue and red
lines, respectively. The absorption at $-3540$ km s$^{-1}$ is unlikely
to be an astrophysical  feature since its $5-6$ \AA~FWHM is lower  than the instrumental resolution.}
\label{fig1}
\end{figure}

H$\alpha$ emission is the only recognizable  spectral feature from
J1659 in the FORS2 data.  For its analysis, each spectrum was rectified through
division by the result of fitting a third-order spline function to the
continuum adjacent to the emission line after excluding nearby
atmospheric absorption bands. Further, the normalized data were
averaged with different weights to maximize the signal-to-noise ratio
of the resulting sum, which is shown in Figure \ref{fig1}.  Given that
the double-peaked morphology of the emission line is resolved, we can
employ two methods of estimating orbital parameters that make use of
the H$\alpha$ line properties during quiescence. The first of these
methods is based on the linear correspondence between the line profile
full-width at half maximum (FWHM) and $K_2$ found for dynamically
studied X-ray transients \citep{2015ApJ...808...80C}. The second method
rests on the existent interdependence also observed for X-ray transients between the ratio of the line double-peak separation $DP$ and
FWHM with $q$ \citep{2016ApJ...822...99C}. The mathematical
formulation of these relations is given below while results from their
application are presented in
\citet{2015MNRAS.454.3351Z,2015MNRAS.454.2199M,2019MNRAS.487.2296T}. As
done in these early studies, the effect of the instrumental broadening
was taken into account when calculating the line parameters. For this, the
model fit to the data was degraded to the instrumental resolution
measured in Section \ref{sec:obs}.

\begin{table}
\caption{The H$\alpha$ emission line parameters. FWHMs have been calculated fitting a Gaussian
  model. the centroid radial velocity (RV) and
peak-to-peak separation ($\Delta V^{pp}$) are derived from a
2-Gaussian model. DP is the double-peak separation measured when both
Gaussian components are forced to have equal height and FWHM. All
units in km s$^{-1}$, except EW (\AA).} 
\label{table1}
\begin{center}
\begin{tabular}{lclc}
\hline                                     & & &                                         \\                
Mean profile                 & & &                                         \\               
\hline
FWHM:              & $3256 \pm 304 $ &$DP$:  & $1916 \pm 76$    \\       
Centroid:            &$ 28 \pm 40$    &FWZI:  & $\gtrsim 3700$    \\        
$\Delta V^{pp}$:  & $1921  \pm 81$  &EW: &  $-130 \pm 10$ \\        
                                   &         & &              \\                
Individual profiles       &         & &              \\     
\hline
FWHM:              & $3164 \pm 284$     & &     \\     
                                   &     & &                                    \\                
  \hline
\end{tabular}		      
\end{center}
\end{table}

Following \citet{2015ApJ...808...80C} we fitted the
individual line profiles with a single Gaussian function using data points within $\pm 10000$ km
s$^{-1}$ with respect to the H$\alpha$ rest wavelength. We derive
from the fits a rounded-off mean FWHM of 3200 km s$^{-1}$ with standard deviation 300 km s$^{-1}$. Applying the
linear correspondence $K{_2} = (0.233 \pm 0.013) \times {\rm FWHM}$, we
obtain $K{_2}=750 \pm 80$ km s$^{-1}$. 
In this calculation we have assumed that the
  FWHM-$K{_2}$ correlation can be extrapolated 
from the upper FWHM value of 2850 km s$^{-1}$ \citep{2015ApJ...808...80C}
to a FWHM of 3200 km s$^{-1}$ as obtained for J1659. Next, following the procedure given in \citet{2016ApJ...822...99C}, we performed
fits to the averaged
H$\alpha$ profile using single and 2-Gaussian models, the latter
having both components with equal FWHM
and height. The fit is illustrated in Figure \ref{fig1} and the
resulting line FWHM and $DP$ are reported in Table \ref{table1}.  To constrain $q$ we make use of the empirical correlation
between this binary parameter and the ratio $DP$ to FWHM given
by $\log q = -(23.2 \pm 2.0) \log (DP/{\rm FWHM}) -(6.88 \pm 0.52)$. As above, we used rounded-off values:
$3200 \pm 300$ km s$^{-1}$  and $1920 \pm 80$  km s$^{-1}$ for the FWHM
and $DP$, respectively. The evaluation of $q$ was computed through  Monte
Carlo randomization where $DP/{\rm FWHM} =0.600 \pm 0.062$ was treated as
being normally distributed about their measured value with standard
deviation equal to its uncertainty. This yielded a non-normal
distribution with median $q=0.018$ and 68\% confidence intervals for
$q$ contained within $0.002-0.23$. The weak
constraint on $q$ is
caused by the $9\%$ uncertainty in $DP/{\rm FWHM}$. 

For completeness and to permit comparison with previous spectroscopy
of J1659 and other X-ray transients, we also  fit a 2-Gaussian model with all parameters
being free. Such a fit has been traditionally  employed to calculate
the peak-to-peak separation (${\Delta V^{pp}}$) and the line centroid
radial velocity. Table~\ref{table1} provides the results from these fits. One can see that 
the values for $DP$ and ${\Delta  V^{pp}}$  are fully consistent, indicating a profile  in accordance with that expected from an axisymmetric disc. 
Table~\ref{table1} also includes the measurements of the line equivalent width (EW) and a
full-width at zero intensity (FWZI). The latter is given as a lower
limit since the ability to establish the extension of the line wings
is set by the noise in the continuum.

The only optical spectroscopy of J1659 published prior to this work
is discussed in \citet{2012ApJ...746L..23K}. It consists of two
consecutive 10 min X-shooter spectra  taken on MJD $55464.9854 -
55465.0000$,  ${\sim}15.5$ h after its initial discovery at $\gamma$-ray
energies and $0.07-0.22$ orbital cycles behind an X-ray dip episode. The
X-shooter spectra were rich in Balmer, Paschen and Helium emission
lines with H$\alpha$ having  ${\rm FWHM} = 1567\pm 28$ km  s$^{-1}$ and ${\Delta
  V^{pp}}=918 \pm 53$ km s$^{-1}$. These values are  a factor 2
smaller  than observed during quiescence (Table \ref{table1}),
reflecting to some degree a decrease  in
the velocity of the  emitting regions due to their outwards
expansion during the outburst event. 
From fig. 1 in \citet{2012ApJ...746L..23K} we infer a line profile
redshift of ${\sim}200$ km s$^{-1}$ (Kaur et al. measure a mean
$156 \pm 42$ km s$^{-1}$ shift from the strongest lines), while in our
data the line centroid is (within the errors) at its rest
position. The amplitude of this velocity drift cannot be due to the
motion of the compact object at the time of the observation
($=q K{_2}\leq50$ km s$^{-1}$ for an adopted $q<0.07$) and is unlikely
to be caused by a precessing
accretion disc since it is too early in the outburst for the growth of
disc eccentricity (see next Sections).  In addition, the H$\alpha$,
H$\beta$ and He {\sc ii} $\lambda 4686$ line profiles were mostly
flat-topped rather than double-peaked. This profile morphology, a
measured ${\Delta V^{pp}} = 1.2 \times K_2$ and the observed velocity
shift depart from the simple picture of emission emerging from
material in Keplerian motion for which
${\Delta V^{pp}} \sim 2\times V(R^{disc}_{out}) > 2 \times K_2$. In
fact, they may be considered spectroscopic evidence for the presence
of an outflow during outburst as observed in other BH transients
\citep{2020MNRAS.tmp.2346C,2019ApJ...879L...4M}.

\subsection{Optical modulations  and their origin}
\label{sec:lcs}
Combined time-resolved i'-,r'- and R-band photometry of J1659 obtained
during five nights in a month interval at the time of the 2011 X-ray
and optical rebrightening that followed the end of the main outburst
is presented in \citet{2018MNRAS.475.1036C}. The time-series
photometry revealed a $2.4149 \pm 0.0006$ h periodicity (referred
hereafter as $P_{ph}^{2011}$) in agreement, within the errors, with
the $2.414 \pm 0.005$ h period found from X-ray dips (henceforth
$P_{dip}^{2010}$). The phase-folded light curve showed a
quasi-sinusoidal modulation with single maximum and ${\sim}0.024$ mag
amplitude. This shape together with the coincident values of
$P_{ph}^{2011}$ and $P_{ph}^{2011}$ was used to support X-ray
irradiation of the donor star as a possible cause for the modulated
variability.  On the other hand, and only noticed in
\citet{2013A&A...552A..32K}, a double-wave modulation with a
$2.41584 \pm 0.00024$ h periodicity ($P_{ph}^{2010}$) was identified
in preliminary R-band light curves taken during the initial stage of
the outburst \citep{2010fym..confP...4K}\footnote{The poster
  presenting this result and light curves is available at
  \url{https://www.researchgate.net/publication/252514067_Multiband_optical_monitoring_of_GRB100925AMAXI_J1659-152}. Note
  that only a fraction of the 2.4 h orbit was observed per night and
  at high airmass given the object visibility. Although the latter
  could affect the quality of the photometry, comparison of the
  nightly multi-band light curves does not show clear trends caused by
  colour-dependent atmospheric extinction.}.  This double-wave shape
is not in line with the sinusoidal wavefront expected for simple X-ray
irradiation of the donor star for which a single period of minimum and
maximum light occurs at inferior and superior conjunction of the
donor, respectively.  In this regard, in the ${{\sim}}2 $ h light
curve presented in \citet{2010fym..confP...4K}, minimum light is
observed around MJD 55466.4568, ${{\sim}}0.70$ orbital cycles after an
X-ray dip. The resulting phase for the minimum adds circumstantial
support against an X-ray heated donor since the likely cause of the
X-ray dips (the bulge) precedes the donor star by a small fraction of
the orbit (e.g., \citealt{2018MNRAS.477.5358R,2019ApJ...882L..21T})
and therefore minimum light should have closely followed after the
X-ray dip event if irradiation were significant. 

Consistent with the above scenario of a weakly or non-irradiated donor
star is the lack of clear narrow line components from Bowen blend
transitions in the X-shooter spectra obtained $0.07-0.22$ orbital
cycles after an X-ray dip (\citealt{2012ApJ...746L..23K}). These narrow features are observed to
emerge from the irradiated donor in neutron star low-mass X-ray
binaries \citep{2002ApJ...568..273S,2003ApJ...590.1041C}.  Instead, the X-shooter data
revealed broad (${\rm FWHM} = 2600$ km s$^{-1}$) Bowen emission with a
velocity offset of $-$775 km s$^{-1}$ with respect to
$\lambda4640$. Interestingly, the radial velocity is
similar to $K_2$. Regarding the latter as a lower limit to the outer disc rim
velocity, a potential explanation for the Bowen blend properties is
that, after the X-ray dip, irradiated inner disc rim regions with
motion pointing towards the observer became visible while regions with
receding velocities were occulted by the structure causing the dip
phenomena.  Finally, let's consider the 0.5-10 keV unabsorbed fluxes
of J1659 at or near the time of the 2010 double-wave and 2011
single-wave optical variability. These were ${\sim}2.2 \times 10^{-8}$
erg cm$^{-2}$ s$^{-1}$ and $(9.4 \pm 0.1) \times 10^{-12}$
erg cm$^{-2}$ s$^{-1}$, respectively
\citep{2011ApJ...736...22K,2012MNRAS.423.3308J}. These
quantities show that major heating of the donor star during the 2011
rebrightening is unlikely since the X-ray flux was a factor ${\sim}2000$ lower
than at the time of the optical modulation found by
\citet{2010fym..confP...4K}. Thus, the observations provide evidence
that an irradiated donor star alone cannot produce the optical
modulations. We will show in Section \ref{sec:irrsh} that X-ray
illumination of the stellar companion is low due to the cover provided
by the vertical extent of the disc. Given that the optical variability
is ascribed to the accretion flow and locked closely to the binary
orbit, this either reflects
long-lasting asymmetries in the disc
structure/emission, their partial occultation by the donor star,
superhumps, or a combination of these effects.

Superhumps are photometric modulations first reported in the optical light curves
of SU UMa dwarf novae during their  large amplitude outbursts (called
superoutbursts) and later recognized in BH X-ray transients
\citep{1996MNRAS.282..191O}. Commonly, superhumps have a single-peaked (hump)  morphology with periods typically a few
percent longer than the orbital period. They are produced when the
outer disc reaches the 3:2 resonance radius ($r_{3:2}$) associated to the 3:1
commensurability between the rotating accreted material and passage of
the donor star (see e. g. \citealt{1991MNRAS.249...25W}). Once $r_{3:2}$  is reached, the
tidal field of the donor star is able to induce the growth of disc
eccentricity and set it in prograde precession
in the apsidal plane. A subgroup of SU UMa dwarf novae,
the WZ Sge stars, are characterized by the occurrence of double-wave
superhumps prior to  the emergence of ordinary superhumps, the former having periods very close (${\sim}0.05-0.1\%$)  to
the binary orbital period (for a review see \citealt{2015PASJ...67..108K}).
%In some cases, WZ Sge stars show post-superoutburst rebrightenings. 
It is mostly accepted that  early superhumps are also caused
by tidal disturbance of the disc by the donor star, developing when 
another resonance, the 2:1, dominates over the 3:2 perturbation.  
This is possible for systems with $q\le0.08$ since then the  radius
of the 2:1 resonance ($r_{2:1}$)  lies within the maximum (tidal
truncation) radius achievable by the accretion disc. This condition is
fulfilled by WZ Sge-type dwarf novae as they usually have $q$ of
$0.06-0.08$. It has been proposed that when $r_{2:1}$ is reached by the
expanding accretion disc during the onset of the superoutburst, the 2:1 resonance
produces  a two-armed  vertically extended pattern in the disc that
causes the double-wave modulation in the light curve \citep{2002PASJ...54L..11K,2012PASJ...64...92U}. 
Furthermore, during this initial phase of the superoutburst, the 2:1
resonance becomes effective in inhibiting the excitation of disc
eccentricity driven by the tidal 3:2 resonance
\citep{1991ApJ...381..268L,1991ApJ...381..259L}. Subsequently, after
sufficient matter removal from the outer disc regions, the disc radius shrinks below
$r_{2:1}$ and is then when ordinary superhumps grow (see
e.g. \citealt{2002A&A...383..574O}).  

How does J1659 fits in this picture?  With estimated $q<0.07$ \citep{2011ApJ...736...22K,2013A&A...552A..32K} its disc has access to both
$r_{3:2}$ and $r_{2:1}$. The expanding disc could  have hit these resonance radii
during the onset of the energetic outburst - this reached an amplitude of 11.5 mag in the V-band \citep{2011ApJ...736...22K}. The apparent
double-wave R-band variability identified by \citet{2010fym..confP...4K} had a
${\sim}0.05$ mag amplitude and was
present during at least the first 18 days after the discovery of  the
transient. Both characteristics are compatible with the properties of
early superhumps in WZ Sge systems with the majority showing
amplitudes $<0.05$ mag (although they can reach 0.35 mag; \citealt{2015PASJ...67..108K}) and  observed
durations of up to ${\sim} 15$ day (Table 9 in \citealt{2013PASJ...65..117N}).  
With regards to ordinary superhumps, their shape is generally
described as saw-tooth, but it can significantly differ as the outburst evolves
\citep{2009PASJ...61S.395K,2008ApJ...681.1458Z}. They can
become at times sinusoidal-like wavefronts as observed in the BH
transient XTE J1118+480
\citep{2002PASJ...54..285U,2002MNRAS.333..791Z} and in the neutron
star systems discussed in Section \ref{sec:irrsh}.  

The relatively low
precision  of the periodicities measured at X-ray and optical
wavelengths prevent us from probing the superhump scenario for J1659 by evaluating the fractional period excess $\epsilon = 1 -
P_{sh}/P_{orb}$  expected for superhumps.
Nonetheless, we constrain it to $\epsilon  <
0.004$ ($2\sigma$) and $<0.006$ ($3\sigma$) by assuming they developed
in the 2011 rebrightening and adopting
$P_{orb}=P_{dip}^{2010}$ and a superhump period
$P_{sh}=P_{ph}^{2011}$. This limit is in line with  the $\epsilon=0.0035$ measured
for the superhumps of XTE J1118+480 ($q=0.024$; \citealt{2002PASJ...54..285U,2002MNRAS.333..791Z}). In the superhump scenario, we can establish a limit on $q$ by utilizing empirical relations between this
parameter and $\epsilon$. Employing $\epsilon \approx 0.25 q
(1+q)^{-\frac{1}{2}} 0.8^{\frac{3}{2}}$ \citep {1992PASJ...44L..15M},
we obtain $q < 0.023$ ($2\sigma$)  and $< 0.034$ at $3\sigma$. Fully
consistent limits are also derived  by applying $q–-\epsilon$ relations
established from a sample of superhumpers that includes not only SU
UMa CVs, but also XTE J1118+480
\citep{2005PASP..117.1204P,2006MNRAS.373..484K,2009PASJ...61S.395K}.  
However, our result must be considered with caution since 
$q–-\epsilon$  relations depend on the superhump stage
\citep{2013PASJ...65..115K,2019MNRAS.486.5535M} and at present they are  not well calibrated at very low mass ratios.

\section{Discussion}
\label{sec:dis}

\subsection{Compact object mass, binary mass ratio and orbital inclination}
\label{sec:mass}
From the FWHM of the H$\alpha$ emission line  detected
in the optical spectrum of J1659 during quiescence we have derived
$K_2 = 750 \pm 80$ km s$^{-1}$ (Section \ref{sec:ha}). This
value in combination with $P_{orb}=P_{dip}^{2010} = 2.414 \pm 0.005$ h
yields a mass function $f(M) = 4.4 \pm 1.4~M_\odot$ (1$\sigma$). This
lower limit to the compact object mass exceeds the range of
observationally determined neutron star masses
\citep{2018MNRAS.478.1377A}, although it is consistent with the
theoretical maximum for a stable rotating neutron star (3.2 $M_\odot$;
\citealt{1974PhRvL..32..324R}). Firmly confirming the BH nature of the
compact object in J1659 requires further constraints on its dynamical
mass which in turn implies knowledge of $q$ and the binary
inclination.  

Beginning with $q$, in Section \ref{sec:ha} we have
established from the analysis of the H$\alpha$ emission line profile
a median value and a 68\% confidence interval of 0.018 and
$0.002-0.23$, respectively. We can put a restriction of
$\lesssim  0.0334-0.067$ from $M{_1} > f(M) = 3.0-5.8~M{_\odot}$ and
by adopting $M{_2} {\sim} 0.20~M_\odot$ for the Roche lobe filling donor star
(a discussion on  the donor star mass and evolutionary stage can be found in \citealt{2011ApJ...736...22K, 2013A&A...552A..32K}). A lower limit of $\gtrsim
0.018$ is obtained for a $<11~M{_\odot}$ BH (see below). Finally,  assuming that the
superhump scenario  for the 2010-2011 outburst optical light curves is
correct,  an upper limit of $q<0.034$ (3$\sigma$) is estimated
employing the limit on the period excess (Section \ref{sec:lcs}). From
these calculations, we constrain $q$ to be in the
range $0.018-0.067$. The uncertainty on the actual value of $q$ has little
effect on the calculated BH mass compared to that already introduced in the
mass function by the statistical error in $K_2$. For instance,  M${_1} \sin^{-3} i =
4.56\pm1.45~M_\odot$  and M${_1} \sin^{-3} i = 4.70\pm1.50~M_\odot$ when considering
$q=0.018$ and $q=0.034$, respectively. We will adopt for the remainder
of this paper  $q=0.034$.

Continuing with the binary inclination, this can be constrained
following the geometrically and physically motivated model for the light
curves of X-ray binaries \citep{1987A&A...178..137F}. Differently to
previous work \citep{2011ApJ...736...22K, 2013A&A...552A..32K}, we take into account that in the particular case of J1659 the disc
outer rim occults the central X-ray source from the donor preventing
X-ray eclipses (see next Section). Thus,  the inclination is bound to
$60^\circ \lesssim i \lesssim 80^\circ$ from the  presence of X-ray
dips and the fact that J1659 did not manifest as an accretion disc
corona source (the disc thermal component was visible in the X-ray
spectrum). Additional information on $i$ can be inferred  from one
attribute  of the H$\alpha$  line profile: the depth of the   central
depression delimited by the two line peaks, which is  near the
continuum level (see Figure \ref{fig1}). This deep absorption core is
expected in accretion discs seen at high inclination \citep{1986MNRAS.218..761H} and they have been reported for a number of  quiescent
eclipsing CVs such as WZ Sge ($i
\approx 77^\circ$; \citealt{1998MNRAS.299..768S}). More recently this
characteristic feature has been detected in the average line profiles of the BH transients
Swift J1357.2-0933 ($i > 70^\circ$, \citealt{2015MNRAS.450.4292T,2015MNRAS.454.2199M}) and
MAXI J1305-704 ($i \approx 75^\circ$; Mata-S\'anchez et al. in prep).  We take from
this comparison that the accretion disc in  J1659
most likely has an inclination $> 70^\circ$ and conclude that the
binary inclination falls in the range $70^\circ \lesssim i \lesssim
80^\circ$. Note that at these inclinations,  partial occultation of
the disc outer structures by the donor star are possible if the former
extended to $r_{3:2}$,  but the resulting shallow dips will be
diluted  in the phase-folded optical light curves by disc
variability. 

The limits on the inclination in combination with the mass function and adopted $q$ yield $5.7 \pm
1.8~M{_\odot}$ ($i=70^\circ$) and $4.9 \pm 1.6~M{_\odot}$ ($i=80^\circ$), implying a
1$\sigma$ interval for the BH mass of:
\begin{align*}
 3.3 &\lesssim M_1(M_\odot) \lesssim 7.5  
\end{align*}
\noindent and a 3$\sigma$ upper limit of
$11.1~M{_\odot}$\footnote{Disregarding the lower limit on $i$ obtained
  from the H$\alpha$ line profile, this becomes $\gtrsim60^\circ$
  leading to upper limits on $M_1$ of $9.5 M{_\odot}$ (1$\sigma$) and
  $14.1 M{_\odot}$ (3$\sigma$).}. The latter rules out the
$20 \pm 3~M{_\odot}$ BH mass predicted in \citet{2011arXiv1103.0531S}
by employing a correlation between X-ray spectral index and QPO
frequency found for BH transients in outburst. The discrepancy is
explained by the fact that the QPO frequency measured for J1659 was
scaled to that observed in the X-ray transient GX 339-4, for which a
$12.3 \pm 1.4~M{_\odot}$ BH (derived from X-ray modeling) was
adopted. Employing instead for GX339-4 the dynamically established
range of possible BH mass
($2.3-9.5~M{_\odot}$; \citealt{2017ApJ...846..132H}), we update to
$3.7-15~M{_\odot}$ the BH mass inferred for J1659 with the above
technique.  Our results also serve to exclude masses
of $16.2-22.7~M{_\odot}$ obtained from
modelling the thermal component of the X-ray spectrum and by assuming a
maximum spinning BH. The above values
represent a revision of those reported in
\citet{2011ApJ...736...22K} and were calculated following the
provisions in their section 4, except for the binary inclination that
was set to $i=70^{\circ}-80^{\circ}$. Estimates for the BH mass were
also obtained by \citet{2012PASJ...64...32Y} from further detailed
spectral and temporal analysis of X-ray data. Using their equation 5
in combination with their restriction on the distance towards the
source and $i=70^{\circ}-80^{\circ}$, the allowed mass range for a
non-rotating BH is $4.3-6.0~M{_\odot}$ ($d=5.3$ kpc) and
$7.0-9.8~M{_\odot}$ ($d=8.6$ kpc) while for a maximum spinning BH
these ranges are a factor 6 larger than for the static case. If
anything, our constraints on the dynamical BH mass in combination
  with the equations in \citet{2012PASJ...64...32Y}  suggest that the  innermost
  stable circular orbit (ISCO) can be consistent either with that
  expected for a non-spinning BH ($r_{\sc{ISCO}}=6 r{_g}$, $r{_g}= GM/c^2$)
or a BH in retrograde rotation ($6< r_{\sc{ISCO}}/r{_g}\leq9$). The latter is
supported by the spectral analysis presented in
\citet{2020ApJ...888L..30R}.  Finally, our limits are in line with the ranges of $4.7-7.8~M{_\odot}$ predicted for the BH by
analyzing the spectral and QPO frequency evolution during the outburst
using two-component advective flow and propagating oscillatory shock
models \citep{2016MNRAS.460.3163M,2015ApJ...803...59D}.

\subsection{Moderately heated mass  donors and visibility of superhumps in low-$q$ accretion-fed X-ray binaries} 
\label{sec:irrsh}
In Section \ref{sec:lcs} we provided a
detailed description of the optical
data obtained for J1659 during eruption
to find that X-ray irradiation of the
donor star may not be the cause for the modulated optical variability
detected in the active phase. To substantiate this conclusion, we
examine here the extent to which the accretion disc casts a shadow on
the donor star surface. Theory and observations have established a disc flare angle of
${\sim}10{^\circ}-20^\circ$ above the orbital plane for active X-ray binaries (\citealt{1982A&A...106...34M, 1996A&A...314..484D,2018MNRAS.474.4717J} and
references therein).  On the other hand,  the open angle subtended by the donor
star as seen by the central X-ray source is $\tan^{-1}(R_2/a)$ with
$R_2/a$ the ratio between its Roche-lobe effective radius and binary
separation. By expressing $R_2/a$ as a function of $q$ \citep{1983ApJ...268..368E},
our limits on $q$ for J1659 of $0.02-0.07$ imply $\tan^{-1}(R_2/a)\approx7{^\circ}-10^\circ$
and thereby that most likely during outburst the disc completely
shielded the donor star from direct exposure to the central X-ray source. From
 this simple geometrical prescription, we expect for BH transients with
$q\lesssim0.07$ a notable reduction in the donor star heating by the
central X-ray source since the angle subtended by the donor at the
compact object is smaller than or comparable to
the lower $10^\circ$ disc flare angle during outburst. Radiation scattered
in the disc corona or X-ray driven wind should also have its heating effect
on the mass donor decreased due to screening by the outer disc
rim. This is
consistent with the BH transient GRS 1915+105 that has
remained at a high X-ray luminosity  since discovery.  In this
system with $q=0.042\pm0.024$ and $P_{orb}=33.85 \pm 0.16$ day,  the
outburst donor star radial velocity curve is not affected by heating
\citep{2013ApJ...768..185S} while the $30.8 \pm  0.2$ day photometric
variability reported in \citet{2007ApJ...657..409N} is inconsistent
with a proposed irradiated donor since it is sub-orbital in nature. On
the other hand, heating of the stellar component was clearly observed during
the outburst of  the BH transient  GRO J1655-40 ($q=0.42\pm0.03$,
$P_{orb}=2.6$ day; \citealt{2000MNRAS.314..747S}) and 
confirmed in GRO J0422+32 ($q=0.11^{+0.05}_{-0.02}$, $P_{orb}=5.1$ h;
\citealt{1997MNRAS.290..303B}). The outburst optical spectroscopy of
the BH transient GX 339-4 ($q=0.18\pm0.05$) revealed  N{\sc iii} Bowen emission
modulated with the 1.76 day orbital period. However, it  is unclear if
it originated (in part) on the donor star \citep{2003ApJ...583L..95H,2017ApJ...846..132H}. 

Given that shielding will weaken the photometric modulation ascribed
to a heated donor star \citep{1996A&A...314..484D}, the superhump
variability can be dominant over the former at high orbital inclinations.
This is opposite to what has been proposed  for systems where the donor is exposed to direct X-ray illumination \citep{2001MNRAS.321..475H}.
However, either if they are due to viscous dissipation or changing
disc area, the detection of superhumps in systems with large inclination angle
such as accretion disc corona systems
($i\gtrapprox80^\circ$) could be difficult due to the foreshortening
of the disc and the potential presence of other variability that
become prominent with inclination. In this regard, 
recurrent optical dips produced in an
equatorial disc outflow characterize the outburst light curve of the
BH transient Swift J1357.2-0933 ($i > 70^\circ$; $q \approx 0.04$;
  $P_{orb}=2.64 \pm0.96$ h) for which  periodic photometric
variability has not been identified yet \citep{2013Sci...339.1048C,2019MNRAS.489L..47C,2019MNRAS.489.3420J}. A  counter example showing the detectability of
superhumps at high inclination is the 1.85 h orbital period persistent
neutron star X-ray binary MS 1603+260, a likely accretion disc corona source
\citep{2003MNRAS.346..684J}. The light curves of its optical
counterpart display both a superhump-like modulation with
$\epsilon=0.0145$ and a jitter behavior in the depth and timing of
partial eclipses by the donor that clearly signal the presence of a
${\sim}5.5$ d disc precession cycle
\citep{2008ApJ...685..428M,2012AJ....144..108M}.  

Accounting for the
fact that $q\lesssim0.08$ for sub-hour orbital period neutron star
binaries (e.g., equation 10 in \citealt{2001MNRAS.321..475H}), the disc will
also shelter the donor star from X-ray heating in systems accreting at
high steady rates. Hence superhumps will be the predominant modulation at
optical bands in ultra-compact X-ray binaries. Support for this is
given by the type-I burster and X-ray dipper source 4U 1915-05 ($P_{orb}=50.0$ min) which displays an
optical modulation with $\epsilon=0.009$ that yields $q\lesssim0.08$
(\citealt{2002MNRAS.330L..37R} and references therein). Superhumps
appear to be the most plausible explanation for the quasi-sinusoidal
wavefront with $\epsilon=0.012$ ($q\approx0.06$) observed in the
ultraviolet for the X-ray dipper 4U 1820-30 in NGC 6624
($P_{orb}=14.1$-min; \citealt{2010ApJ...712..653W}) and we anticipate
this also to be the case for the 17.4 min periodic sinusoidal-like
modulation in the optical light curve of 2S 0918-549
\citep{2011ApJ...729....8Z}.
 
\subsection{Concluding remarks}

We argue in Section \ref{sec:lcs} that the properties of the optical
variability observed in J1659
are consistent with the presence of early superhumps, as those observed
during the superoutbursts of WZ Sge dwarf novae, and that the periodic
variability during the 2011 rebrightening event could be due to
ordinary superhumps. Undeniably, a reliable interpretation of the
optical modulations must be furnished with the publication of a
rigorous analysis of the photometry in \citet{2010fym..confP...4K} and
\citet{2010fym..confP...6O} that may deliver a robust value for
$P_{ph}^{2010}$ and other possible periodicities.
Early superhumps have not been firmly seen in other BH transients in part
due to the fragmentary observations collected for most sources and 
the presence of strong flickering that make difficult the detection of
low amplitude modulations (e.g. \citealt{2006ApJ...651..401H,2006ApJ...644..432Z}). The best evidence for a BH accretion disc reaching the 2:1
resonance radius is the long delay in the growth of ordinary
superhumps witnessed during the 2019 outburst of the $q=0.072$ BH
transient MAXI J1820+070
\citep{2018ATel11756....1P,2020ApJ...893L..37T}. This delay is considered for CV superoutbursts lacking early superhumps as an indirect
evidence for the suppression of disc eccentricity development by the  action of the 2:1 resonance.

\section*{Acknowledgments} \noindent 

First, we thank the invaluable support provided by the administrative
and IT divisions in our institutions that enable us to continue
the research activities, including this work, during the  national
lockdowns due to the COVID-19 pandemic. This work has been supported by the Spanish MINECO under grant
AYA2017-83216-P.  MAPT acknowledge support via a Ram\'on y Cajal
Fellowship RYC-2015-17854.  We thank the anonymous referee
  and Sandeep Rout for useful comments. {\sc iraf} is distributed by the National Optical Astronomy Observatory, which is operated by the Association of Universities for Research in Astronomy (AURA) under a cooperative agreement with the National Science Foundation. 
Tom Marsh is thanked for developing and sharing his package
{\sc molly}.

\section*{DATA AVAILABILITY} \noindent 
The data underlying this article are publicly available
at \url{http://archive.eso.org/cms.html} under program ID 091.D-0865(A)

\bibliographystyle{mnras}
\bibliography{mybib} % if your bibtex file is called example.bib

\end{document}